# Manuscript

**Variability of the Sun's luminosity places constraints on the thermal equilibrium of the convection zone**

L.E.A. Vieira[1], G. Kopp[2], T. Dudok de Wit[3], L. A. da Silva[4,1], F. Carlesso[1], A. Barbosa[1], A. Muralikrishna[1], R. Santos[1]

[1]Instituto Nacional de Pesquisas Espaciais, Av. dos Astronautas, 1758, São José dos Campos, Brazil. e-mail: luis.vieira@inpe.br

[2]University of Colorado, Laboratory for Atmospheric and Space Physics, 3665 Discovery Drive, Boulder, 80302 CO, USA.

[3]LPC2E, CNRS, CNES and University of Orléans, Orléans, France.

[4]National Space Science Center, State Key Laboratory of Space Weather, Chinese Academy of Sciences, Beijing, China

## Summary Paragraph

Luminosity, which is the total amount of radiant energy emitted by an object, is one of the most critical quantities in astrophysics for characterizing stars. Equally important is the temporal evolution of a star's luminosity because of its intimate connection with the stellar energy budget, large-scale convective motion, and heat storage in the stellar interior. The Sun's luminosity and its variation have not been measured to date because current observations of the solar radiative output have been restricted to vantage points near the Earth. Here, we model the solar luminosity by extending a semi-empirical total solar irradiance (TSI) model that uses solar-surface magnetism to reconstruct solar irradiance over the entire $4\pi$ solid angle around the Sun. This model was constrained by comparing its output to the irradiance in the Earth's direction with the measured TSI. Comparing the solar luminosity to the TSI on timescales from days to solar cycles for cycles 23 and 24, we find poor agreement on short timescales (< solar rotation). This is not unexpected due to the Earth-centric viewing geometry and short-term irradiance dependence on surface features on the Earth-facing solar disk. On longer timescales, however, we find good agreement between the luminosity model and the TSI, which suggests that the extrapolation of luminosities to multi-cycle timescales based on TSI reconstructions may be possible. We show that the solar luminosity is not constant but varies in phase with the solar cycle. This variation has an amplitude of 0.14% from minimum to maximum for solar cycle 23. Considering the energetics in the solar convection zone, it is therefore obvious that a steady-state input from the radiative zone at the solar minimum level would lead to a gradual reduction in the energy content in the convection zone over multi-century timescales. We show that the luminosity at the base of the convection zone should be approximately 0.032% higher than that at the solar surface during solar minimum to maintain net energy equilibrium through the solar cycle. These different energy-input scenarios place constraints on the long-term evolution of the total solar irradiance and its impact on the solar forcing of climate variability. These results highlight the convection zone's role as an energy reservoir on solar-cycle timescales and set constraints for dynamo models intending to understand the long-term evolution of the Sun and solar analogs.

**Introduction**

The star of which we have the best knowledge of electromagnetic emission and its variability is the Sun (Spruit, 1977; Hudson, 1988). The total solar irradiance (TSI), which is the spectrally-integrated radiant flux at the top of the Earth's atmosphere and normalized to 1 AU, is essentially the portion of the solar luminosity (or total outgoing radiant energy) in the Earth's heliocentric direction. The TSI currently provides the most indicative direct measurements of the solar luminosity. These measurements have been restricted to regions near the ecliptic plane and span only the last four decades, so are much shorter than the approximately $10^5$-year thermal-relaxation timescale of the convection zone (Mark S. Miesch, 2005; Featherstone and Miesch, 2015). TSI models based on current spaceborne measurements extend this time range back thousands of years via indicators (such as the sunspot number) of solar-surface magnetic-activity variability, which is the main driver of fluctuations in the TSI (Fligge, Solanki and Unruh, 2000; Krivova *et al.*, 2003; Domingo *et al.*, 2009; Shapiro *et al.*, 2011; Vieira *et al.*, 2011; Ball *et al.*, 2012; Yeo, Solanki and Krivova, 2013; Lean *et al.*, 2020).

The main identified mechanisms and associated timescales of TSI variability are solar oscillations (5 minutes), granulation (tens of minutes), sunspot evolution (few days), facular variability (tens of days), the longitudinal asymmetry of the magnetic activity (27 days), and the active network and latitudinal asymmetry (one solar cycle, i.e. approximately 11 years) (Hudson, 1988; Vieira *et al.*, 2012; Kopp, 2016). In this work, we focus on variations of the irradiance on timescales from days to the solar cycle related to the evolution of active regions and the active network. Variations on daily timescales are primarily due to the balance of dark (sunspots) and bright features (faculae and network) on the solar surface and their relative position to the observer. Shorter-term variations of the solar irradiance related to solar flares are also detectable, although these depend strongly on the total amount of energy released and flare location (Woods, Kopp and Chamberlin, 2006; Kretzschmar *et al.*, 2010), but are energetically insignificant compared to the luminosity.

Sunspots cause an easily detectable decrease in the TSI. This decrease occurs because intense magnetic fields within sunspots block turbulent and thermal convection, thus inhibiting upwelling thermal energy transport from deeper layers to the photosphere (Borrero and Ichimoto, 2011). The reduction of the temperature within sunspots causes a reduction of the surface opacity. Surfaces of constant optical depth within sunspot umbrae are located at deeper geometric depths. Additionally, sunspots have lower gas pressure than the surrounding regions and the quiet Sun., a phenomenon first described by A. Wilson in 1769 (Wilson, 1774, 1783). Since optical depth unity depends on the sunspot's relative position on the solar surface to the observer, maximum decreases in the TSI occur when the sunspots are near the center of the disk, causing decreases as large as ~0.3% (Willson *et al.*, 1981; Kopp, Lawrence and Rottman, 2005). The magnetic field configuration also determines the positive irradiance contributions, which are due to bright features (faculae and network). A similar mechanism to that which causes irradiance depletion in sunspots, but on smaller spatial scales, causes enhancements due to these features. Since the facular magnetic flux tubes are narrow, the outflow of radiation from their hot walls exceeds the energy blocked from the geometric optical-depth effects. For bright features, the maximum enhancements occur when observed about 60º from the disk center. Consequently, the dark and bright structures' distributions and geometries lead to a non-isotropic radiation field (Spruit, 1977; Steiner, 2005).

*What problem are we addressing?*

The fundamental question we address is "What is the net energy transport out of the convection zone?" This is determined by whether the luminosity blocked by sunspots is balanced by the increased emissions from bright features when integrated over the entire solar surface. The possibility of a thermal energy gain or loss from the convection zone has received considerable attention (Livingston, 1982; Newkirk, 1983; Spruit, 2000). The convection zone's energy-exchange mechanisms act on the main energy reservoirs, which are kinetic ($E_k$), magnetic ($E_m$), and thermal energy ($E_{th}$). The total energy is conserved only if the net flux through the inner and outer boundaries of two spherical shells immediately encompassing the upper and lower boundaries of the convection zone are equal. The variability of the total energy ($E_{tot}$) thus depends on the net fluxes of the kinetic energy, enthalpy, radiative diffusion, Poynting flux, and viscous energy. Additionally, we must consider the internal and gravitational potential associated with the background pressure stratification.

Considering that the Sun's thermal timescale, the Kelvin-Helmholtz timescale, is approximately 15 million years (Spruit, 1977), we assume that over the 11-year solar-cycle timescale the energy flux coming from the radiative zone and crossing its boundary (the tachocline) with the differentially-rotating outer convective zone, is constant. The kinetic- and magnetic-energy flux through the outermost boundary (the photosphere) can be evaluated from in situ observations of the solar wind plasma density and velocity and the interplanetary magnetic field. Le Chat et al. (2012) estimated that the solar wind energy flux at 1 AU is approximately $8.5 \times 10^{-4}$ Wm$^{-2}$. Additionally, they found that the solar wind energy flux is independent of the solar-wind speed and latitude within 10% and that this quantity varies only weakly over the solar cycle. This energy flux is nearly insignificant compared to the radiative contribution, whose value at 1AU is approximately equal to of the nominal TSI value (1361 Wm$^{-2}$), first established by Kopp and Lean (2011) and subsequently defined by IAU 2015 Resolution B3 (Prša *et al.*, 2016). Thus, the radiative contribution dominates all other energy-loss mechanisms, and any imbalance in the net steady-state input from the solar radiative zone and the outgoing radiative energy from the photosphere should lead to a gradual change in the energy contained in the convection zone over multi-century timescales. Such a long-term energy change, if found, would profoundly impact our understanding of solar variability (Güdel, 2007) and solar forcing of climate (Solanki, Krivova and Haigh, 2013).

Here, we contribute to this issue by providing the first reconstruction of the solar luminosity on timescales from days to solar cycles. We do so by extending a semi-empirical TSI model to estimate the Sun's radiant energy output in 4π steradians, using observations of solar-surface magnetic activity on the Earth-viewable portion of the solar disk and a flux-transport model to estimate that activity in regions that are not viewable from the Earth's vantage point. Integrating the estimated angularly-dependent irradiance over 4π steradians effectively gives a model of the solar luminosity.

## Approach to estimating the irradiance over 4π steradians

We distinguish three terms in this paper (Wilhelm, 2010): (1) "Solar Irradiance" or "Irradiance" is the spectrally-integrated radiant flux at 1 AU for some heliocentric position; (2) "Total Solar Irradiance" (TSI), for consistency with colloquial use, is the spectrally-integrated radiant flux at the top of the Earth's atmosphere and normalized to 1 AU; and, (3) "Luminosity" refers to the net radiative output power from the Sun, being an integration of the solar irradiance over 4π steradians; The definitions (1) and (2) correspond to the radiative flux density given in $W/m^2$ as defined by Parr, Datla and Gardner (2005).

Observations of the solar irradiance from vantage points other the Earth's are currently not available and will not be in the near future. We have measurement access only to the TSI. To overcome this directional limitation, we reconstruct the evolution of the irradiance for any vantage point located at 1 AU from the Sun by using the spatial distribution of photospheric magnetic features on the Earth-facing solar disk. Such models have been shown to be remarkably successful in reproducing total and spectrally-resolved irradiance observations for the last four solar cycles (Krivova *et al.*, 2003; Wenzler *et al.*, 2006; Domingo *et al.*, 2009; Ball *et al.*, 2012). Vieira *et al.* (2012) first employed such models to estimate the irradiance out of the ecliptic plane based on observations made by the Helioseismic and Magnetic Imager (HMI) (Schou *et al.*, 2012) on the Solar Dynamics Observatory (SDO) spacecraft for the ascending phase of Solar Cycle 24. Their conclusions, however, were hampered by the lack of coverage of the far side of the Sun and the poor visibility of polar regions, where magnetic-field measurements are less accurate (Güdel, 2007). Here, we improve upon that approach by using a flux-dispersive assimilation model developed by Schrijver and Derosa (2003), which we shall refer to as Flux Transport Model, to extend the coverage to the full solar surface. This model estimates the evolution of the solar-surface magnetic flux and thereby enables irradiance estimates from any vantage point based on full-surface activity. Using this approach, we estimate the luminosity for Solar Cycles 23 and 24 and, for the first time, obtain a realistic estimate of the luminosity over several solar cycles.

The flux transport model was slightly modified to incorporate new/updated observations and statistical properties of the magnetic field structures observed on the solar surface. While the version by Schrijver and Derosa (2003) was based on data from the Michelson Doppler Imager (MDI) on the Solar and Heliophysics Observatory (SoHO), here we use a modified version that incorporates HMI/SDO data as well, allowing extension of temporal coverage beyond 2010. The modified version includes updates of the HMI calibration to that given by Liu *et al.* (2012) and a modified flux half-life from 5 years to 10 years, which was incorporated to match the HMI/SDO observations.

To compute the irradiance in a desired heliocentric direction, we estimate the distribution of the magnetic concentrations on the solar surface from that vantage point. From these flux-transport-model magnetic-flux maps over the entire solar surface, we estimate the fraction of the solar disk that is covered by the quiet Sun, by sunspots (umbrae and penumbrae), and by bright elements (faculae). We subsequently compute the radiative output of the solar atmosphere by using solar atmosphere models (Wilhelm, 2010) specific for each feature. In this way, the evolution of the density flux at a given wavelength (λ), colatitude (θ), longitude (ϕ), and $\mu(\theta, \phi)$ can be expressed as:

$$F(\lambda, \theta, \phi, \mu, t) = \alpha_u(\mu, t)\, \Delta F_u(\lambda, \mu) + \alpha_p(\mu, t)\, \Delta F_p(\lambda, \mu) + \alpha_f(\mu, t)\, \Delta F_f(\lambda, \mu) \quad (1)$$
$$+ \alpha_{eph}(\mu, t)\, \Delta F_{eph}(\lambda, \mu) + F_q(\lambda, \mu)$$

where $\mu = \mu(\theta, \phi)$ is the cosine of the angle between the normal to the solar surface at some position and the observed line-of-sight. The filling factors for sunspot umbrae and penumbrae are represented by the time-dependent coefficients $\alpha_u(\mu, t)$ and $\alpha_p(\mu, t)$, respectively. The filling factors for the bright elements are indicated by $\alpha_f(\mu, t)$ (faculae) and $\alpha_e(\mu, t)$ (ephemeral regions). The corresponding differences between the time-independent radiative fluxes of the bright and dark components of the model and the quiet-Sun, $F_q(\lambda, \mu)$, are represented by $\Delta F_u(\lambda, \mu)$, $\Delta F_p(\lambda, \mu)$, $\Delta F_f(\lambda, \mu)$, and $\Delta F_{eph}(\lambda, \mu)$.

Instead of employing continuum images to compute the filling factors for sunspot umbrae and penumbrae, we assume that sunspots are in the plasma pressure balance regime with the surrounding regions. In this way, we segment the magnetic pressure distribution using two thresholds: $Th_1$ for sunspot penumbrae; and $Th_2$ for sunspot umbrae.

The filling factors of the individual pixels of the facular elements ($\alpha_f$) were determined by the relationship $\alpha(f) = \min(1, B/B_{sat})$, where B is the magnetic field intensity and the free parameter $B_{sat}$ is the saturation level. Due to the low spatial resolution of the synthesized magnetograms, we cannot properly evaluate the filling factor for ephemeral regions. Assuming that these regions (which occur in the quietest portions of the solar surface) are generated by a process that is not modulated by the global dynamo, we filter out pixels with low intensity magnetic fields by applying a threshold ($B_{eph}$). The contribution from ephemeral regions to solar irradiance is assumed to be constant ($\mathcal{F}_{eph}(t) = \iint \alpha_{eph}(\mu, t)\, \Delta F_{eph}(\lambda, \mu)\, d\lambda d\mu = constant$) and thus appears as an additional free parameter ($\mathcal{F}_{eph}$). We thus are not able to detect long-term variations in the solar irradiance and luminosity, as those are expected to include variations of ephemeral-region contributions.

We effectively model each location on the solar surface by the corresponding modeled spectrum at that position on the solar disk. After summing over all locations and integrating over all wavelengths, we obtain the solar irradiance for any given heliocentric direction.

To estimate the five (5) free parameters (Th$_1$, Th$_2$, B$_{sat}$, B$_{eph}$, and $\mathcal{F}_{eph}$), we compute the difference between the model's outputs and the TSI composite provided by (Dudok de Wit *et al.*, 2017)[1] as well as and the bright (faculae and network) and dark (sunspots) components estimated by Yeo et al. (2013) employing the SATIRE model.

We use a genetic optimization algorithm to estimate the model's free parameters (Metcalfe and Charbonneau, 2003; Vinnakota and Bugenhagen, 2013) that minimize the difference between the modelled and the observed TSI.

Our figure of merit cost function ($\chi$) is defined as the sum of the weighted mean squared error ($MSE$) for the three datasets employed,

$$\chi = \begin{aligned} w_1 * MSE(TSI_{\{obs\}}, TSI_{\{model\}}) + w_2 * MSE(F_{\{fac,SATIRE\}}, F_{\{model\}}) \\ + MSE\left(w_3 * \left(F_{\{u,p,SATIRE\}} - F_{\{u,p,model\}}\right)\right) \end{aligned} \qquad (2)$$

where $MSE$ is the average squared difference between the modeled values ($Y_i$) and the actual observations/estimates ($\hat{Y}_i$). For $N$ observations, we can write

$$MSE(\hat{Y}_i, Y_i) = \frac{1}{N} \sum_{i=1}^{N} (Y_i - \hat{Y}_i)^2 \qquad (3)$$

We reduce outlier effects in the estimate of the free parameters by using the weighting function $w_j$, where the index $j$ refers to the dataset. We define the weighting function, $w_j$, as

$$w_j = 1/(1 + r_j^2), \qquad (4)$$

where

$$r_j = (Y_i - \hat{Y}_i)/(g * s_j * \sqrt{1 - h_j}), \qquad (5)$$

---
[1] The data is available at: https://spot.colorado.edu/~koppg/TSI/TSI_Composite-SIST.txt

and

$$s_j = MAD(Y_i - \hat{Y}_i)/0.6745, \tag{6}$$

where the median absolute deviation ($MAD$) is the residuals from their median. We use the median because it is more robust to outliers; the constant 0.6745 makes the estimate unbiased for the normal distribution. The Hat matrix leverages *($h_j$)* adjust the residuals by reducing the weight of high-leverage data points (Vinnakota and Bugenhagen, 2013). The tune parameter ($g$) is set to 2.385 (Cauchy weight value).

As mentioned, the flux transport model is based on two data sets: (a) From 1996 to 04/2010, the model is based on MDI/SoHO data; and (b) from 05/2010 to 2019, the model is based on HMI/SDO data. The solar images associated with these data sets have different spatial resolutions. To reduce inconsistencies in the detection of active regions, we estimate the free parameters separately for each solar cycle, transitioning between the two instruments shortly after the 2008 solar minimum. By fitting the free parameters for each cycle, we risk obtain a time series that is not smooth at the boundaries of the cycles. Additionally, long-term trends are not detectable.

Table 1. Free parameter estimates for the model.

| Solar Cycle | Period | $Th_1$(G) | $Th_2$(G) | $B_{sat}$(G) | $B_{eph}$(G) | $F_{eph}$ (Wm$^{-2}$) |
|---|---|---|---|---|---|---|
| **SC #23** | 1996-2008 | 377.7 | 758.2 | 399.5 | 1.8900 | 2.0340 |
| **SC #24** | 2009-2016 | 300.4 | 713.1 | 294.3 | 3.0480 | 2.1695 |

## Evolution of the Solar Luminosity

*The effect of active regions on the global radiative field*

In Fig. 1, we illustrate the sunspot darkening effect on the global radiative field for a spherical shell at 1 AU for October 29, 2003. The short-term decrease in the TSI during the passage of groups of sunspots across the solar disk (Fig. 1a) was the largest ever recorded (Fig. 2a, blue line). However, a single-location observation at the Earth's heliographic position does not capture the spatial or temporal extension of the luminosity darkening caused by the active regions' presence during this period. Our reconstruction of the radiative field for a shell at 1 AU shows the vast extension of this darkening over an expansive range of directions covering almost half of the shell's surface (Fig. 1b) and causing a considerable relative decrease in solar luminosity. This large decrease does not match the unusually large depth of the TSI, which occurred because the large sunspot group reached its maximal extent precisely when it was facing the Earth (Fig. 2a, red and blue lines). We point out that because the fitted the datasets separately, the transition at the boundaries of the Solar Cycles 23 and 24 is not smooth.

Comparing the solar luminosity to the TSI on timescales from days to the solar cycle (see Figs. 2a and 2b), we note that single-location Earth-centric observations do not capture the global radiative field's evolution as active regions emerge and decay (see video - lum_2003_v04.mp4), which highlights the importance of considering the luminosity and not just the irradiance to understand the energetics of the convection zone.

The solar luminosity differs from the TSI in important physical ways. While the TSI corresponds to observations at a specific heliographic latitude and longitude tied to the Earth's position, the irradiance captures the full extent of directional inhomogeneities. This effect can be seen in Fig. 3a, which presents the latitudinal distribution of the irradiance for an observer at Earth's heliographic longitude. As suggested by (Knaack *et al.*, 2001), the latitudinal variation occurs because the effects of bright features extend farther toward high latitudes than the depletion caused by sunspots. The effects of the emergence and decay of the sunspots on the irradiance also depend on the heliographic longitude, as shown in Fig. 3b for solar-equatorial observers at the Earth's (Lon: +0º) and three other longitudes. The different longitudes show different phases and amplitudes of irradiance variability due to the as-observed positions of the active regions on the solar disk.

The latitudinal asymmetry of the occurrence of active regions does not only affect the radiative field during major events, as in October 2003, but also the distribution throughout the solar cycle, as shown in Fig. 4a, which displays the latitudinal dependence of the irradiance over the last two solar cycles. The polar-viewed irradiance has nearly the same overall solar-cycle amplitudes as the TSI due to the predominance of faculae when viewed from the poles (Fig. 4b). Note, however, the absence of the abrupt short-duration decreases in the polar-viewed irradiances, as those vantage points are relatively insensitive to the near-equatorial sunspots causing those decreases in the TSI. Although most of the variability occurs near the equatorial region, the latitude at which the maximum is largest occurs is highly dependent on the distribution of the active regions and the solar cycle phase. In particular, the poles' irradiance decreases faster than that at low latitudes during the descending phase (Fig. 4c). This is not unexpected given the latitudinal dependence of the activity through the solar cycle. What is more surprising is the hemispherical asymmetry of the irradiance that is observed in this period. Note that the irradiance peaks in the southern hemisphere after the magnetic-field polarity's reversal for both Solar Cycles 23 and 24. The most striking discrepancy occurs during the descending phase of Solar Cycle 23 when the south pole's irradiance exceeds that of the north pole's. This effect is caused by the different evolution of each hemisphere's magnetic activity. Indeed, during the ascending phase of Solar Cycle 24, the northern hemisphere is more active than the southern hemisphere. During the extended minimum between Solar Cycles 23 and 24, the inverse occurs as the flux is lower at the north pole than at the south pole.

The average profile of the irradiance (see Fig 4b) shows a modulation resulting from the balance between bright and dark features. The average irradiance over the last two solar cycles peaks at mid-latitudes (NH: 31.5°; SH 45.0°), with the southern hemisphere's peak being slightly higher than the northern hemisphere's peak. The variability is also consistently higher and more extensive at low latitudes, with a maximum at the equator (Fig. 4d). We estimate that the standard deviation at the equator is about 0.5 Wm$^{-2}$, while the standard deviation at the poles is about 0.36 Wm$^{-2}$. We point out that this variability range would be detectable by present-day instrumentation such as VIRGO/SoHO and TIM/SORCE, were they observing from those vantage points, so such direct observations would be achievable with present-day capabilities. Analyzing the TSI and solar luminosity time-series employing moving averages in different scales, we find that despite the heliocentric positional differences discussed above, they agree on timescales longer than a few solar-rotational periods. We next discuss these variabilities on solar-cycle and longer timescales.

*Evolution of the solar luminosity for solar-cycle timescales*

We show in Fig. 2b a reconstruction of the solar luminosity ($L_\odot$) for Solar Cycles 23 and 24, while Fig. 2c gives the estimate of the excess of power that is emitted and blocked by bright and dark features, respectively. For both solar cycles, the luminosity increases in phase with the magnetic activity cycles. The luminosity reaches its maximum during Solar Cycle 23 with a level that is approximately 0.14% higher than the minimum between Solar Cycles 22 and 23. This maximum is about twice as large as the maximum observed for Solar Cycle 24. No differences are apparent between the three solar minima covered by the reconstruction within the model's error, which is defined as the difference between the observed and the modeled TSI (see Figure 2d). The yellow line shows the uncertainties estimated for the TSI composite (Dudok de Wit *et al.*, 2017).

In addition to the modulation of the solar spectral emission of the magnetic structures imprinted on the solar surface, luminosity changes are also due to a combination of the magnetic fields' thermal effects in the convection zone. These effects are: (1) the magnetic field reservoir acts as source and sink of thermal energy; and (2) the magnetic field changes the heat transport coefficient. The first effect is related to the generation of magnetic fields that involve converting the energy of motion into magnetic energy. As advection in the convection zone is thermally driven, this effectively converts thermal into magnetic energy. The decay of magnetic structures has the opposite effect. The second effect comes from the suppression of convective motion by magnetic fields, which leads to a reduction in heat transport efficiency in the convection zone.

The net effect of the magnetic fields' thermal and kinetic effects in the convection zone can explain the solar cycle systematic variations of the meridional flow at the solar surface. Hathaway and Upton (2014) found that the meridional flow speed is fast at cycle minima and slow at cycle maxima.

**Evolution of the thermal energy content in the convection zone**

We find that on timescales from days to years the power blocked by sunspots does not compensate for the output power enhancement due to bright features. On timescales of one solar cycle, the excess of radiative emissions that is due to the increase in emissivity of small-scale magnetic-field regions should increase the surface's cooling rate. Consequently, the cooling of the surface should eventually lead to a slight reduction of the temperature in the convection zone. However, the characteristic timescales on which thermal perturbations propagate in the convection zone are depth-dependent. We can estimate this thermal timescale as a function of depth from

$$\tau(z) \equiv U(z)/L \approx \frac{1}{L} \int_{R-z}^{R} 4\pi r^2 u \, dr, \tag{7}$$

where $U(z)$ is the energy in the layer between a given depth (z) and the surface, and $u$ is the thermal energy per unit volume. From this we conclude that the thermal timescales should vary as the cube of depth below the solar surface. As a result, the thermal response timescale of the Sun at 20 Mm depth, for example, is about 11 years, while at the base of the convection zone the timescale is about $10^5$ years. We note that the 20 Mm depth corresponding with 11-year energetic timescales is typical of the depths of conveyor belt flows causing meridional circulation. Because of this cubic dependence on depth, near-surface disturbances such as solar-surface magnetic-activity regions have a much larger impact on solar-cycle timescales than deeper ones.

According to Miesch (2005), if the convection zone is in thermal equilibrium, then the energy fluxes should balance such that

$$\langle F_r^{KE} + F_r^{EN} + F_r^{RD} + F_r^{PF} + F_r^{VD} \rangle_{\theta\phi t} = \frac{\langle L_\odot \rangle_t}{4\pi R_\odot^2}, \tag{8}$$

where the terms $F_r^{KE}$ and $F_r^{EN}$ represent the kinetic energy and enthalpy flux by convective motions, respectively. The radiative diffusion is represented by $F_r^{RD}$, while $F_r^{PF}$ indicates the Poynting flux. Finally, the viscous energy flux is indicated by $F_r^{VD}$. The brackets indicate an average over the surface and a specified time window, $t$. The energy flux through the convection zone is small relative to the internal energy of the plasma, so we expect equilibrium to occur on relatively long timescales (Fan, 2004; Mark S. Miesch, 2005).

Recently, Christensen-Dalsgaard (2021) presented an overview of the current understating of the solar structure and evolution, including a detailed description of the modeling of the physical processes. The author noted that the variation of the solar irradiance in phase with the solar cycle of around 0.1% peak to peak leads to a difficulty to estimate the appropriate luminosity corresponding to equilibrium conditions. To address this point, we point out that the reconstruction of the luminosity presented in the previous section allows us to explore the evolution of the thermal energy in the convection zone for Solar Cycles 23 and 24. To compute the thermal energy in the convection zone, we can assume that its variation is given by the difference between the luminosity at the inner and outer shell boundaries, that is

$$\frac{dE_{\text{th}}(t)}{dt} = L_{RZ}^{RD}(t) - L_{1AU}^{RD}(t) \tag{9}$$

where $L_{RZ}^{RD}(t) = \iint F_r^{RD}(\theta,\phi,t)d\theta d\phi$ is the luminosity at an inner shell, while $L_{1AU}^{RD}(t) = L_\odot(t) = \iint F_{r,1AU}^{RD}(\theta,\phi,t)d\theta d\phi$ is the luminosity at the outer shell, at 1 AU (as no energy is lost between the outer surface of the convection zone and a shell at 1 AU).

While we can assess the variability of the luminosity for the external shell based on observations, the luminosity for the internal shell is not measured. Let us therefore consider two different scenarios for this lower-boundary input (see Table 2). Being primarily interested in luminosity timescales of solar cycles to millennia, which are much smaller than the $10^5$-year thermal-relaxation timescale of the entire convection zone, we can treat this input as being constant on those shorter timescales.

Table 2: Energy-input scenarios

| Scenario | Luminosity at an inner shell | Description | Value |
|---|---|---|---|
| #1 | $L_{1AU,Ref}^{RD}$ | Average value for the solar minimum between solar cycles 23 and 24. | Constant |
| #2 | $L_{Eq}^{RD}$ | Luminosity for which the system returns to equilibrium over the solar cycle. | Estimated by optimization. |

In the first scenario, we assume that the irradiance at the inner shell is uniform with a constant value that is equal to the average value for the solar minimum between solar cycles 23 and 24, ($L_{1AU,Ref}^{RD}$). In this way, we can write Equation (9) as

$$\frac{dE_{\text{th}}(t)}{dt} = L_{1AU,Ref}^{RD} - L_{1AU}^{RD}(t) \tag{10}$$

Integrating Eq. (10) over the time period 1996/07 to 2019/12, we estimate that the convection zone's thermal energy would decrease with time as shown in Fig. 5b (red line). Consequently, this minimal level of luminosity associated with the minimum state of the magnetic activity would not lead to an equilibrium state on the 11-year solar-cycle timescale, as expected from simple energetics. In this scenario of continuing energy loss from the convective zone, a cooling of the near-surface layer should occur, eventually reaching equilibrium at a lower surface temperature than present but with a timescale substantially longer than the 40-year observation records of the magnetic activity and TSI.

From Eq. 7, solar-cycle timescales involve thermal changes at depths between the surface and 24 Mm, which corresponds to a timescale of approximately 22 years, so the decrease in surface temperature corresponding to the energy loss rate in this scenario would be approximately 0.5 K over the period presented here. This change in the temperature is inconsistent with observations of the solar irradiance, which suggests that this scenario can be ruled out. We note that this depth is contained in the region between surface and the base of the Sun's surface shear layer, which is about 60 Mm below the surface, where the equatorward return flow of the meridional circulation seems to occur.

In our second energy-input scenario, we assume that the system returns to equilibrium over the solar cycle. Assuming that the irradiance at the inner shell is uniform and constant, we need to estimate the value for which the variation of convection zone's energy content is zero after the solar cycle. In this way, we can write Equation (9) as

$$\frac{dE_{\text{th}}(t)}{dt} = L_{Eq}^{RD} - L_{1AU}^{RD}(t) \tag{11}$$

The value of $L_{Eq}^{RD}$ can be estimated by requiring that the thermal energy returns to the initial value at the end of the solar cycle. This condition requires that the luminosity be about 0.032% higher than the average for the solar-cycle minimum 23-24 (Fig. 5b; blue line). The convection zone's energy content would increase in each cycle for values of the inner shell's luminosity greater than this quasi-equilibrium condition (Fig. 5). Generalizing this "equilibrium" scenario, the convection zone evolves in an energy cycle modulated by the magnetic activity. In the initial portion of each solar cycle after the terminator, which is thought to indicate the end of the prior cycle (McIntosh *et al.*, 2014), the thermal energy in the near-surface depths of the convection zone decreases as the presence of the bright features allows enhanced outgoing radiant energy. After the maximum of the activity cycle, the thermal energy continues to decrease. In the descending phase of the cycle, with fewer active regions and the consequent reduction of the excess emission, the input energy flux from the radiative zone exceeds the convection zone's losses and the net energy content returns to the level observed during the minimum of the activity. More specifically, based on our luminosity model, for this "equilibrium" scenario, which we based solely on solar cycle 23, we find that the convection zone's thermal energy decreases just after the terminator and reaches a minimum at this cycle's descending phase. As the luminosity falls below the threshold level ($L_{Eq}^{RD}$), the energy content increases until it returns to the level at the previous minimum. The same pattern repeats for cycle 24, although the thermal energy in the near-surface layers would be expected to end slightly higher since this was a much weaker cycle than the one upon which $L_{Eq}^{RD}$ was determined.

While we determined the value of $L_{Eq}^{RD}$ based only on solar cycle 23, this cycle is fairly representative of the average sunspot activity levels for the last 300 years and thus represents normal solar activity since the end of the Maunder Minimum. As the input energy to the convective zone is assumed to be constant and as this region's thermal time constant is $10^5$ years, the estimated input energy should be based on the mean over a similarly long time period. No such directly-measured record exists of solar activity, however, so we instead consider using the mean of the sunspot number – the longest direct observational record of solar activity and the basis for most historical reconstructions – as being representative of long-term solar activity. The mean sunspot number since 1700 differs from that during solar cycle 23 by only 0.5 %, and thus basing $L_{Eq}^{RD}$ on this cycle alone is a reasonable representation of the long-term activity level.

The generalization of the energy cycle through solar cycle 23 described above could approximately describe the evolution of the energetics if no secular trends on the modulation of the luminosity are assumed. This scenario is consistent with the suggestion by (Schrijver *et al.*, 2011) that the solar-surface magnetic field that was measured during the deep 2008-2009 minimum, when the solar magnetic activity decreased to a comparable level found in the quietest areas between active regions sustained by small magnetic bipolar ephemeral regions, may provide the best estimate of the conditions that prevailed during the Maunder Minimum.

In contrast to this view, several reconstructions of the solar-surface magnetic field that are based on sunspot records (Solanki, Schüssler and Fligge, 2000) and cosmogenic isotopes (Solanki *et al.*, 2004) suggest the existence of secular trends in its evolution. Additional support for such trends comes from semi-empirical reconstructions of the TSI (Krivova, Vieira and Solanki, 2010; Vieira *et al.*, 2011; Coddington *et al.*, 2016; Wu *et al.*, 2018). Although these models correctly incorporate the processes that lead to the variability of the TSI, they make no assumptions regarding the energetics in the convection zone. One of the key messages of our study is that these near-surface energetics may place constraints on long-term luminosity variations of the TSI.

Figure 6a shows yearly averaged reconstructions of the historical TSI and the modeled luminosity ($L/4\pi r^2$). We show two TSI reconstructions: (1) the TSI/SATIRE model (Krivova, Vieira and Solanki, 2010) (blue); and (2) the TSI/NRLTSI2 model (Coddington *et al.*, 2016) (red). The differences in these models' reconstructions arise from their assumptions regarding the emergence of ephemeral regions. For the Maunder Minimum, the SATIRE model suggests that the level of the irradiance would be about 0.7 W m$^{-2}$ lower than that occurring during the Solar Cycle 23-24 minimum. Assuming that the solar radius remained constant over this time ($R_\odot=6.957\times10^8$ m) (Prša *et al.*, 2016), this would imply an increase of the solar-surface temperature by approximately 0.73 K from 1700 to the Solar Cycle 23-24 minimum. Based on the TSI/NRLTSI2 model this temperature increase would be approximately 0.5 K.

We present in Fig. 5b an estimate of the variability of the near-surface thermal energy (depth = 24.2 Mm; $\tau$ = 22 years) assuming that the irradiance at the inner boundary of the convection zone is constant at the level of the Maunder Minimum. We note that under this assumption, the near-surface thermal energy content would have decreased by 0.6 to 0.8% since 1700. This result suggests that in addition to the long-term modulation of the irradiance, changes in the energy flux input at the base of the convection zone would be necessary to account for the model's long-term trend. A steady increase of the energy flux at the base of the convection zone would be required to maintain thermal equilibrium over this 400-year time range. This differs from the two scenarios that we presented but is within reason given the uncertainties of the long-term changes in the TSI over this time span.

Long-term reconstructions are important because the TSI is the main energy input to the Earth's climate system (Hansen *et al.*, 2005; Schmidt *et al.*, 2006; Jungclaus *et al.*, 2010), exceeding all other inputs by more than three orders of magnitude. For that reason, even a minute change of 1‰ in the long-term value of the TSI would have a major impact on climate. Not surprisingly, the potential existence of multi-decadal trends in the TSI has been examined with great care (Douglass and Clader, 2002; Jungclaus *et al.*, 2010; Lean, 2010; Ermolli *et al.*, 2013; Kopp, 2014). Particular attention has been given to the change in TSI estimated since the Maunder Minimum, although the uncertainties are comparable to the changes over this 400-year time range.

To maintain long-term thermal equilibrium, our reconstructions of the luminosity indicate that the net energy transport out of the convection zone varies in phase with the solar cycle. This result is consistent with historical TSI reconstructions, as their possible small secular variations would cause surface-temperature changes based on our thermal-energy model that have not been directly observed at the accuracy levels needed over these timescales, but would, in all likelihood, be measurable with current space-based instrumentation. On solar-cycle timescales, we find that the amount of luminosity that is blocked by sunspots is not immediately balanced by increased emissions associated with bright features. To reach a steady state over solar-cycle timescales, the average amount of energy entering the convection zone at the tachocline does not correspond to the luminosity level during recent solar minima but rather to a cycle average that is 0.032% greater. Longer-term changes in the luminosity may be possible but would be driven by deeper layers of the convection zone.

The anisotropy of the solar irradiance highlights the importance of expressing the energy budget in terms of luminosity and not just the total solar irradiance, which corresponds to a single vantage point. This anisotropy stresses the need for measuring the solar irradiance from vantage points outside of the ecliptic plane.

Our analysis does not confirm or dismiss the hypothesis that a long-term trend in solar forcing is present since the end of the Maunder Minimum. However, the modeling of such long-term trends in the TSI should be consistent with the energy budget of the convection zone. We encourage such thermal-energetic constraints be included to provide more consistent long-term irradiance reconstructions.

# Figures

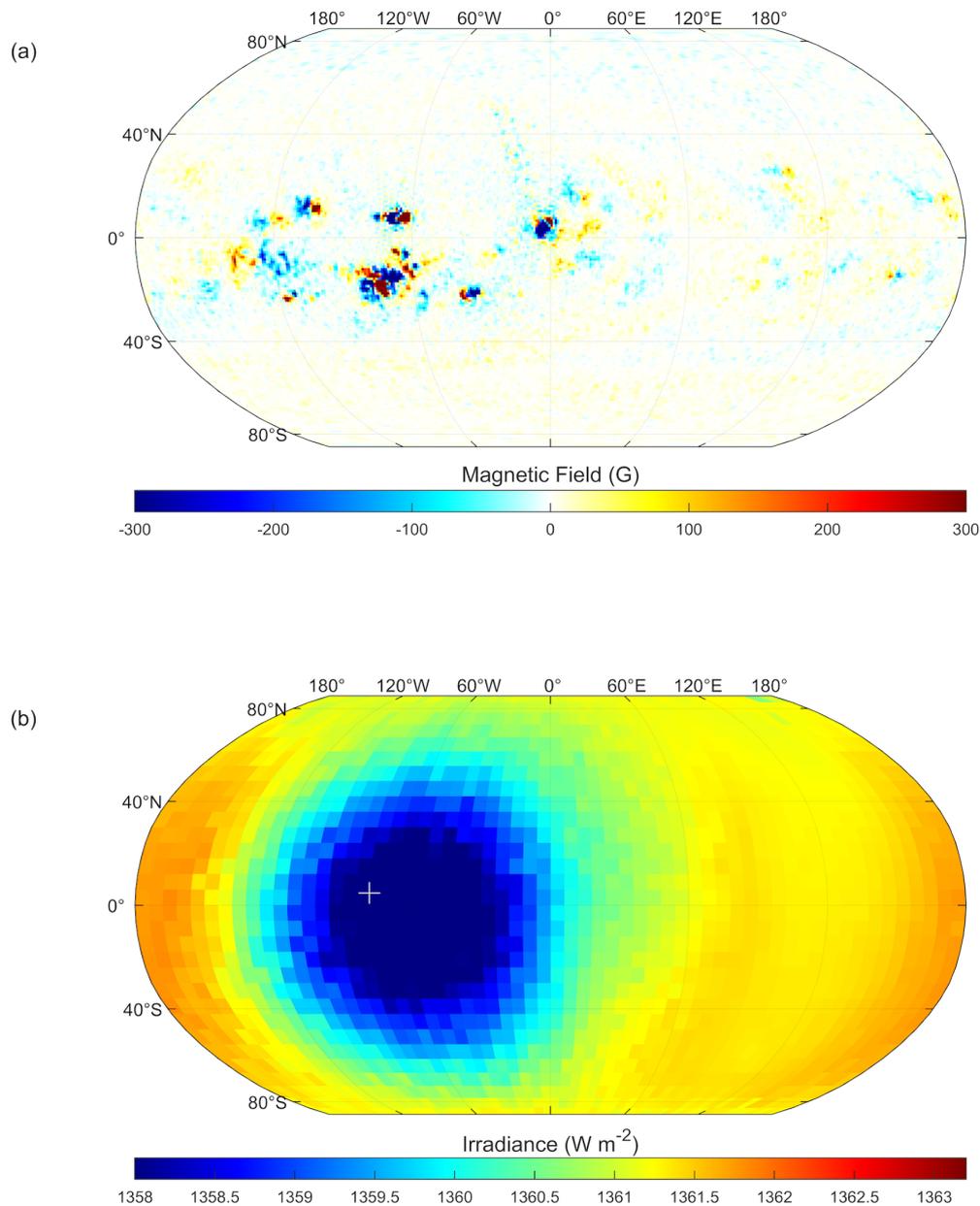

**Figure 1**. Reconstruction of the irradiance by extending TSI models to 4π steradians using a flux-transport model to estimate magnetic-activity positions and magnitude variations. (a) Distribution of the magnetic field concentrations on the solar surface for October 29, 2003, when huge sunspot groups were facing Earth. (b) Reconstruction of the solar irradiance as a function of heliocentric location. The white cross in panel (b) indicates the Earth's approximate heliographic latitude and longitude. An animation of this figure is available. Panels (a) and (b)

are shown on the lower right and upper left portions of the animation, respectively. The right side of the animation shows the luminosity time series similar to Figure 2. The animation begins on January 2nd, 2003 and ends on December 31st, 2003. The realtime duration of the animation is 73 seconds.

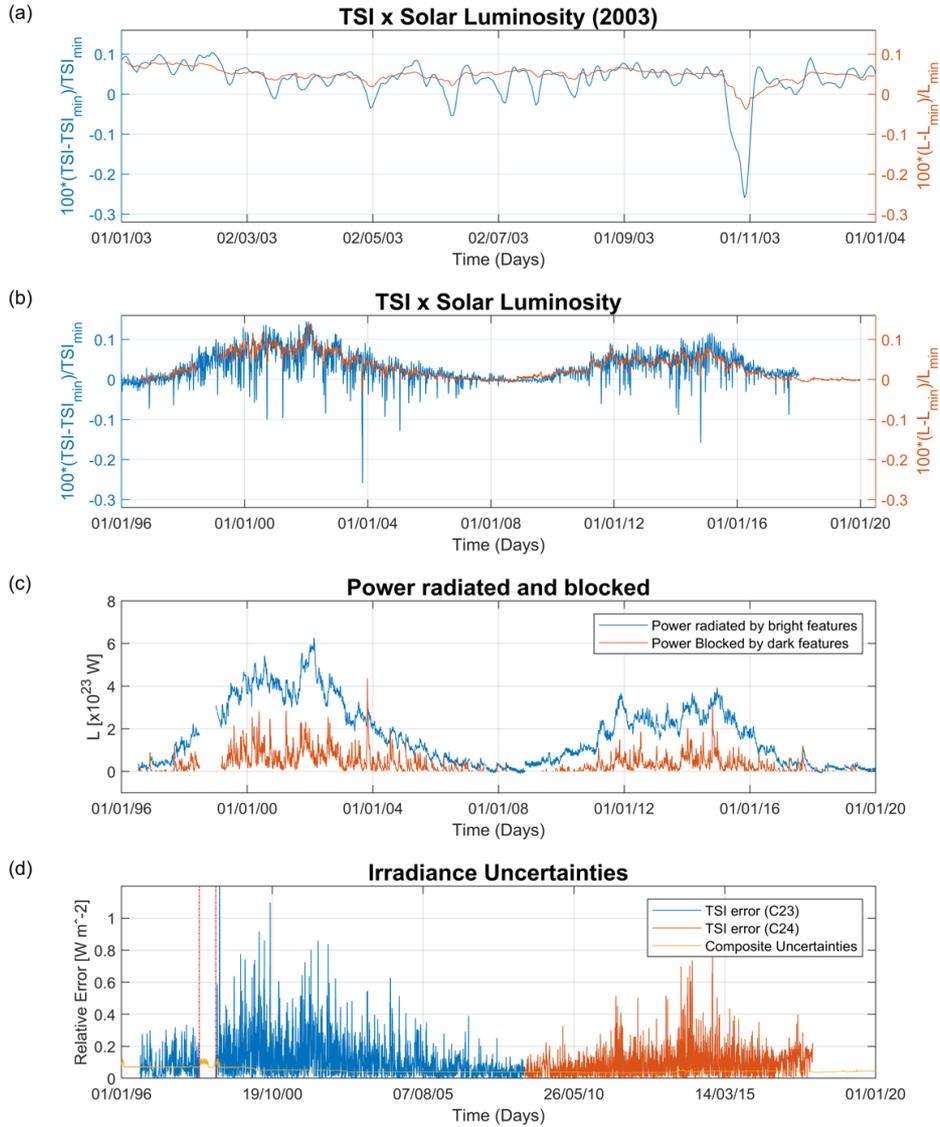

**Figure 2**: Reconstruction of the solar luminosity for cycles 23 and 24. Panels (a) and (b) present comparisons between the percentage variation of the total solar irradiance observations (blue) and the reconstructed luminosity (red) for 2003 and for solar cycles 23/24, respectively. The reference level for both quantities is the average value for the year 2008 during the solar minimum between cycles 23 and 24. (c) Power enhancement due to bright features (blue) and deficit due to power blocked by sunspots (red). (d) Difference between the TSI composite and our model for each cycle (blue/red for 23/24). The estimated TSI composite uncertainties are shown in yellow. The dashed red lines indicate the 3-month period during which the SoHO

spacecraft lost contact (beginning June 24, 1998). The tick labels for the time are in the format dd/mm/yy.

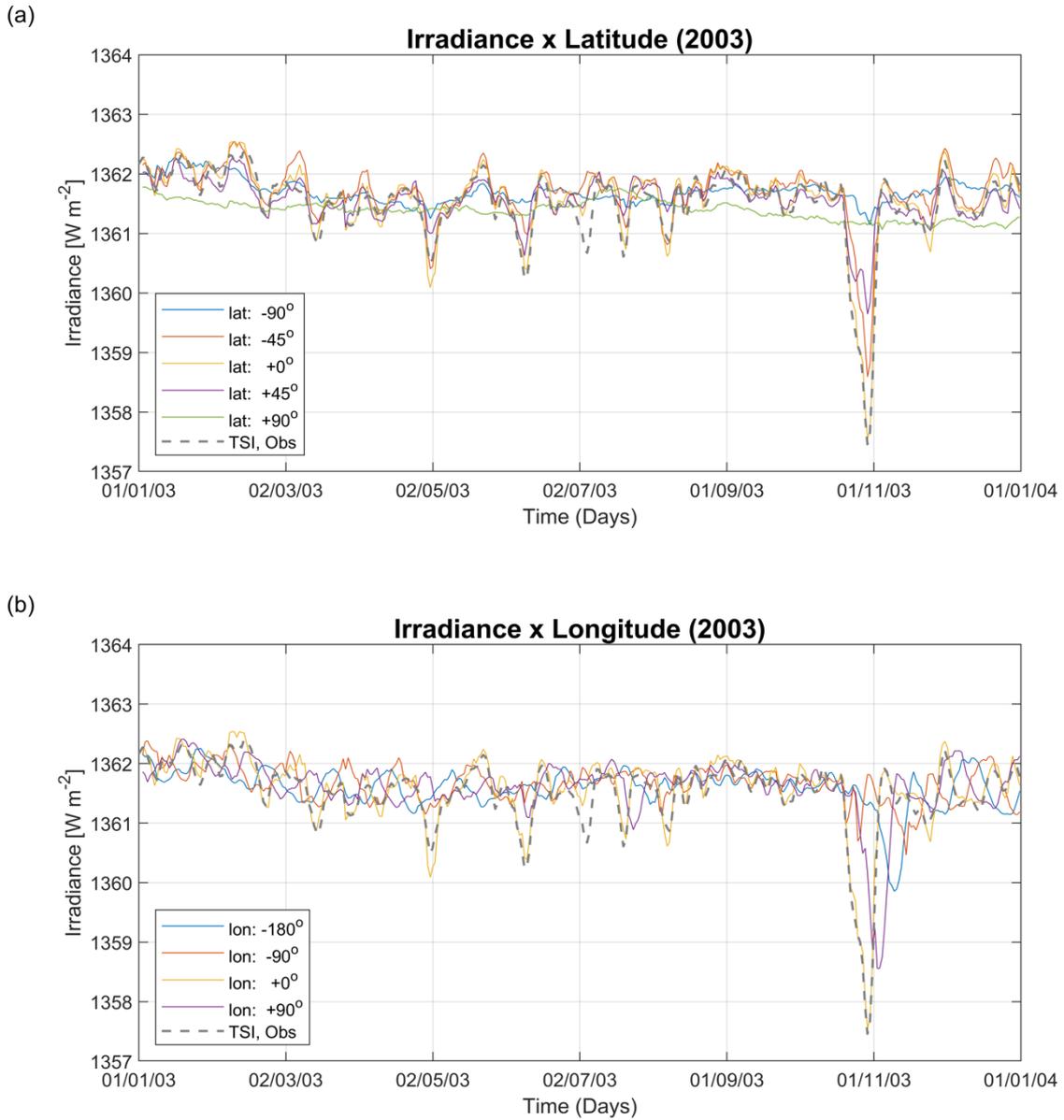

**Figure 3.** Reconstruction of the irradiance on the heliographic latitude for an observer at Earth's heliographic longitude (a) and dependence on heliographic longitude for an observer at the solar equator (b) for 2003. For reference, we show the TSI observations as dashed lines in both panels. The tick labels for the time are in the format dd/mm/yy.

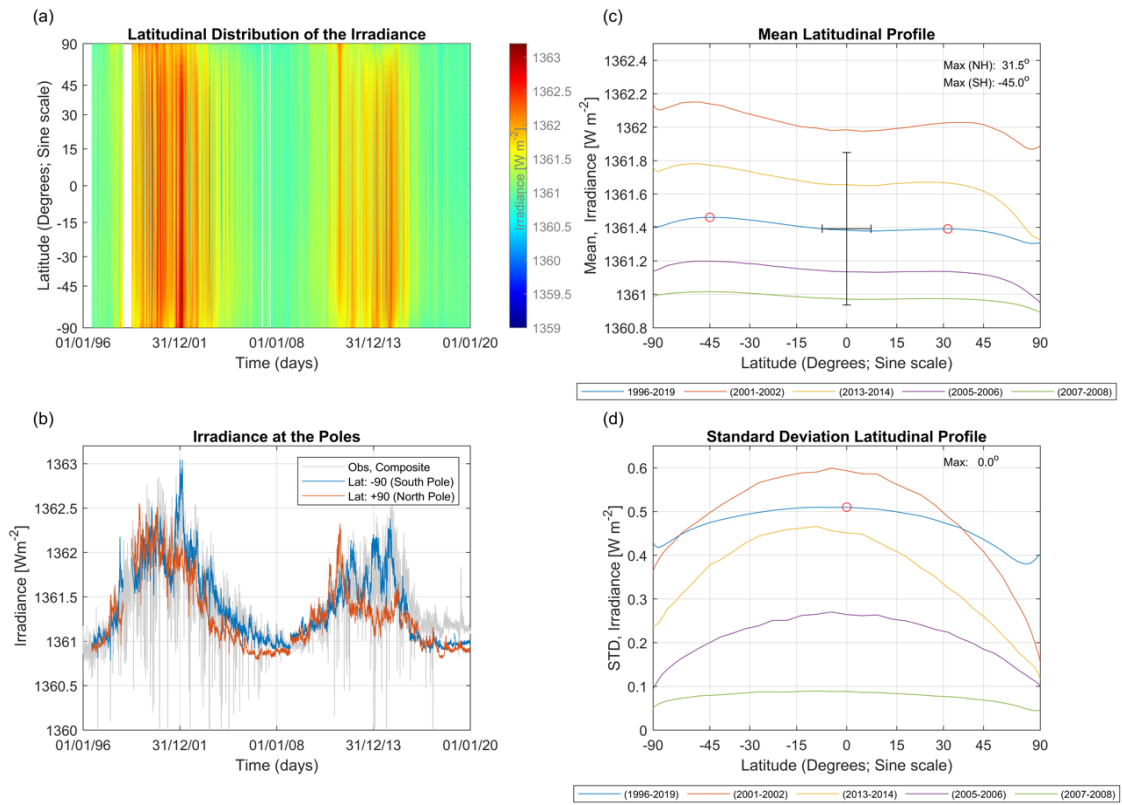

**Figure 4**: The latitudinal dependence of the irradiance over the last two solar cycles. (a) Latitudinal distribution of the irradiance for an observer at Earth's heliographic longitude. (b) The TSI composite (gray line) and the solar irradiance viewed from the heliographic south (blue line; model) and north poles (red line; model). (c) Average latitudinal irradiance profile for the whole period (blue line), at solar maximum (2001-2002 and 2013-2014), during the descending phase (2005-2006), and at solar minimum (2007-2008). The latitude of maximum for each hemisphere for the whole period average is indicated in the figure (red circles). (d) Latitudinal standard deviation profile for the same periods in the upper panel. The red circle indicates the maximum for the average profile for the whole period. The tick labels for the time for panels (a) and (b) are in the format dd/mm/yy.

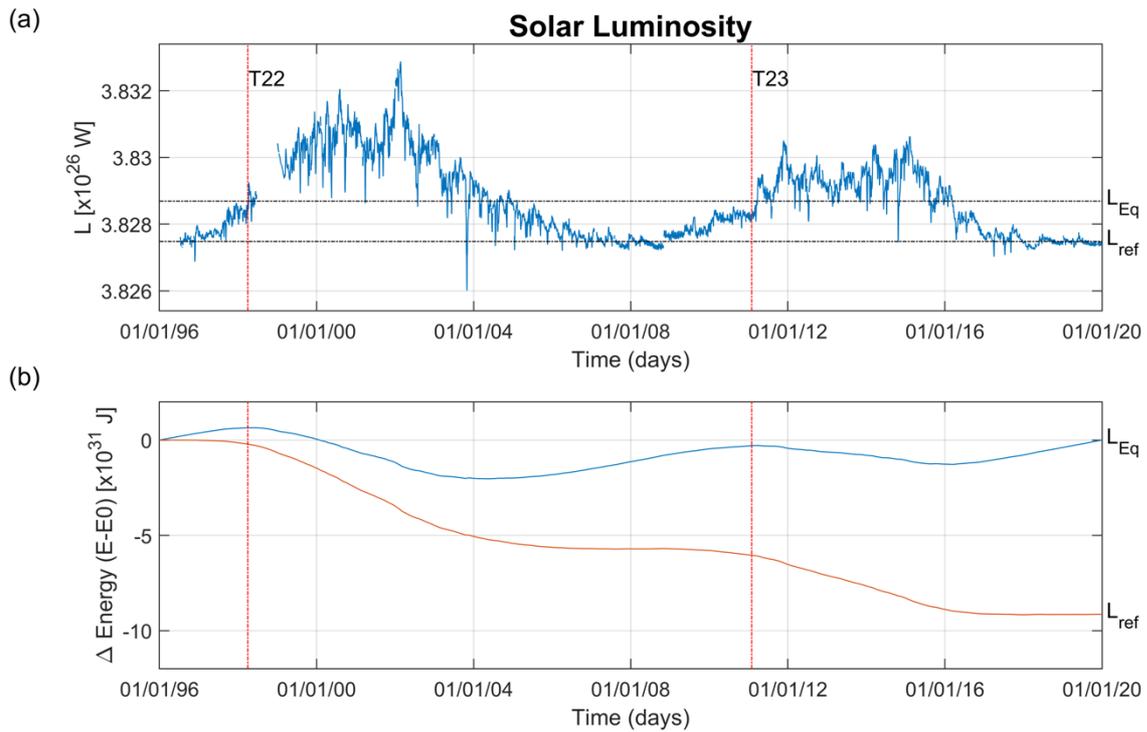

**Figure 5.** Energetics in the solar convection zone for cycles 23 and 24. (a) Evolution of the luminosity. The vertical red lines T22 and T23 indicate the terminator for cycles 22 and 23, respectively, as estimated by McIntosh et al. (2020). These terminators supposedly indicate the completion of the corresponding numbered solar cycle. (b) Variation of the thermal energy for two steady-state energy-input scenarios from the radiative zone: (1) Constant at solar minimum level (L_ref; red); and (2) average that would maintain convection-zone thermal equilibrium over solar cycles for 23 and 24 (L_eq; blue). The time tick labels are in the format dd/mm/yy.

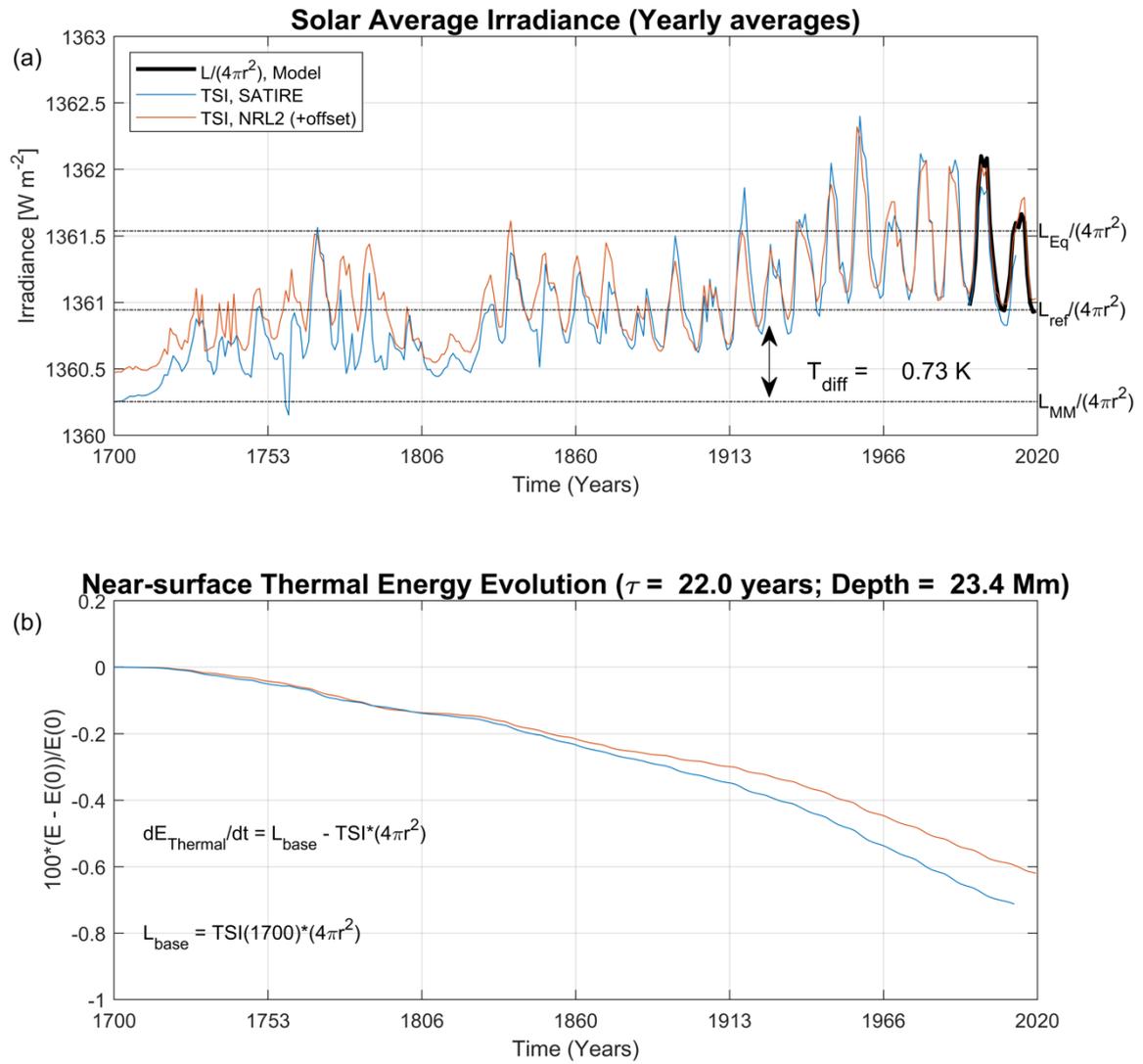

**Figure 6.** Convection-zone thermal energetics place constraints on the variability of solar luminosity on solar-cycle to multi-century timescales. (a) Yearly average irradiance reconstructions for the TSI SATIRE and NRLTSI2 models are shown in blue and red, respectively. The recent solar-cycle minimum level ($L_{ref}/(4\pi r^2)$), the average that would maintain convection-zone thermal equilibrium over solar cycle 23 ($L_{eq}/(4\pi r^2)$), and the level at the Maunder Minimum for the TSI/SATIRE model are indicated by the three horizontal lines in the figure. (b) The corresponding thermal energy evolution for each TSI model for depths in the convection zone between the surface and 24.2 Mm for an input-energy scenario in which the energy at the base of the convection zone is constant and equal to the level of that at the

Maunder Minimum. Note that in this figure the solar radius is assumed to be constant $(r = R_\odot)$.

**Acknowledgements**

- **Funding:**
    - L.E.A.V.: Brazilian Space Agency (AEB) for the funding (TED n. 004/2020-AEB; PO 20VB.0009).
    - G.K.: NASA SIST NNX15AI51G
    - T.D.W.: CNES
    - L.A.S.: Financial support from China-Brazil Joint Laboratory for Space Weather
    - F.C.: PIBIC/CNPq for the grant n. **300274/2022-0**
    - A.B.: FAPESP 2019/13181-0

- **Data**
    - The authors thanks Karel Schrijver and Mark de Rosa for the flux transport model runs employed to estimate the evolution of the solar-surface magnetic flux. LMSAL Evolving Surface-Flux Assimilation Model: Version 2. The model is available at: https://www.lmsal.com/forecast/.